\documentclass[toc]{PoS}

\usepackage{ulem}
\usepackage{amssymb}
\usepackage{amsmath}
\usepackage{fontenc}
\usepackage{times}
\usepackage{mathptmx}
\usepackage{natbib}

\def\lcdm{$\Lambda$CDM}

\def\omm{\Omega_{\rm 0,m}}
\def\omdm{\Omega_{\rm 0,dm}}
\def\omb{\Omega_{\rm 0,b}}
\def\omlambda{\Omega_{\rm 0,\Lambda}}

\def\vbx{v_{{\rm b}x}}

\def\zsun{\rm{~Z_{\odot}}}

\def\msun{\rm{~M_{\odot}}}

\def\apj{ApJ}

\def\apjl{ApJ}
\def\mnras{MNRAS}

\def\aj{AJ}

\def\nar{New Astronomy Reviews}
\def\jcap{J. of Cosm. and Astrop. Phys.}

\def\nat{Nature}

\def\apss{Ap\&SS}	

\def\prd{Phys. Rev. D}

\title{Bulk Flows and End of the Dark Ages with the SKA}

\ShortTitle{Bulk Flows and End of the Dark Ages with the SKA}

\author{
  \speaker{Umberto~Maio}$^{1,2}$,
  Benedetta~Ciardi$^3$,
  Leon~V.~E.~Koopmans$^4$\\
  {}$^1$ INAF -- Osservatorio Astronomico di Trieste, via G. Tiepolo 11, Trieste, Italy; Marie Curie Fellow, E-mail: \email{maio@oats.inaf.it}\\
  {}$^2$Leibniz Institute for Astrophysics (AIP), An der Sternwarte 16, D-14482 Potsdam, Germany, E-mail: \email{umaio@aip.de}\\
  {}$^3$ Max Planck Institute for Astrophysics, Karl-Schwarzshild-Stra{\ss}e 1, Garching b. Muenchen, Germany; E-mail: \email{ciardi@mpa-garching.mpg.de}\\
  {}$^4$ Kapteyn Astronomical Institute of the University of Groeningen, Landleven 12, Groeningen, The Netherlands; E-mail: \email{koopmans@astro.rug.nl}\\
 }

\abstract{
The early Universe is a precious probe of the birth of primordial objects, first star formation events and consequent production of photons and heavy elements.
Higher-order corrections to the cosmological linear perturbation theory predicts the formation of coherent supersonic gaseous streaming motions at decoupling time.
These bulk flows impact the gas cooling process and determine a cascade effect on the whole baryon evolution.
By analytical estimates and N-body hydrodynamical chemistry numerical simulations including atomic and molecular evolution, gas cooling, star formation, feedback effects and metal spreading for individual species from different stellar populations according to the proper yields and lifetimes, we discuss the role of these primordial bulk flows at the end of the dark ages and their detectable impacts during the first Gyr in view of the upcoming SKA mission.
Early bulk flows can inhibit molecular gas cooling capabilities, suppressing star formation, metal spreading and the abundance of small primordial galaxies in the infant Universe.
This can determine a delay in the re-ionization process and in the heating of neutral hydrogen making the observable HI signal during cosmic evolution patchier and noisier.
The planned SKA mission will represent a major advance over existing instruments, since it will be able to probe the effects on HI 21-cm at $z\sim 6-20$ and on molecular line emissions from first collapsing sites at $z\sim 20 - 40$.
Therefore, it will be optimal to address the effects of primordial streaming motions on early baryon evolution and to give constraints on structure formation in the first Gyr.
}

\FullConference{
  Advancing Astrophysics with the Square Kilometre Array\\
  Giardini di Naxos, Sicily, Italy\\
  June 8 - 13, 2014
}

\begin{document}

\section{Introduction}

\noindent
The initial phases of the Universe are extremely interesting for modern Astrophysics, because they are supposed to witness the birth of the first stars and galaxies.
Features and details of the mechanisms leading to this event are still unknown and the properties of primordial generations are largely debated \cite[][]{Maio2011b, Biffi2013, deSouza2013, deSouza2014, Wise2014}.
Primordial star formation processes determine the production of photons in different energy bands, marking the end of the dark ages and the start of the cosmic dawn.
Ongoing feedback from stellar evolution impacts the following baryon history and cause the occurrence of the first heavy elements in the Universe.
Such processes are crucial in the build-up of small primordial visible objects, such as gamma-ray-burst host galaxies \cite[][]{Campisi2011, Salvaterra2013, Ghirlanda2013, Barkov2014}, gaseous damped Lyman-alpha absorption systems \cite[][]{Simcoe2012, Maio2013} and very early faint galaxies \cite[][]{Jeeson2012, Dunlop2013, Ouchi2013, MaioViel2014}.
\\
In addition, the emission by atomic \cite[][]{Wouthuysen1952, Field1958, Bahcall1969} and molecular \cite[][]{Shchekinov1986, Shchekinov1991, Ciardi2001, Kamaya2002} lines in the pristine gas would be strongly dependent on the thermal state of the high-redshift ($z$) inter-galactic medium.
This whole picture is affected by the properties of the initial perturbations that can leave their imprint in the infancy of primordial structures.
Indeed, supersonic streaming motions are expected to be originated as a result of the relative velocities between dark matter and baryons following recombination, at $z\simeq 10^3$ \cite[][]{Tseliakhovich2010}.
They are due to non-linear corrections to cosmological perturbation theory that predicts coherent gas motions on Mpc scales with typical velocities with a rms value of $\sim 30\,\rm km/s$.
These bulk flows could be able to influence cosmic gas emission \cite[][]{McQuinn2012, Pritchard2012} and star formation events at $z\sim 10-30$ \cite[][]{Maio2011a, Stacy2011, Greif2011, MaioIannuzzi2011, Fialkov2014}, with possible consequences on the kinetic SZ effect and on the determination of the temperature-density equation of state \cite[][]{Dalal2010}.
Although direct observations are still very challenging and outside the possibilities of current facilities, they could be probed by the future Square Kilometre Array (SKA) \cite[][]{SKA2004, SKA2013} that will detect radiative emissions by cosmic gas at redshift $z\sim 6-30$.
At those epochs the Universe is largely pristine, dominated by H atoms that can form H$_2$ and HD and lead to gas cooling and collapse.
Molecular-rich gas in dense regions is likely to emit through collisional excitations around $\sim$~533~GHz and such features can be used to identify the very first star forming sites.
The following star formation processes would locally inject large amounts of entropy \cite[e.g.][]{Maio2013entropy} and would produce energetic photons that will boost HI 21-cm (1.4~GHz) emission via Ly-$\alpha$ pumping, allowing the gas to be seen against the cosmic microwave background \cite[][]{Wouthuysen1952, Field1958, Madau1997,WyitheLoeb2003, Ciardi2001}.
The resulting temperature contrast in the optically thin limit depends on the optical depth, $\tau$, which is a function of the neutral H fraction.
Therefore, 21-cm analysis is a very suitable probe of the ages before re-ionization and can help place constraints on cosmic star formation, early galaxy evolution and on the implications of primordial bulk flows \cite[][]{Dalal2010,McQuinn2012,Visbal2012,Mellema2013}.
Given the fact that SKA will observe in different frequency ranges -- between 50 MHz (SKA1-LOW) and 14~GHz (SKA1-MID) in Phase 1 and up to 24~GHz at 10 times larger sensitivity in Phase 2 -- these radiative emissions will result within its instrumentational potential and, if detected, will represent a strong advancement for our theoretical understanding of primordial structures.
\\
In the following, the impacts of baryon streaming motions for early gas emissions and structure formation will be outlined and the possibilities for the different frequency ranges of SKA will be discussed in more detail.

\section{Streaming motions at the of the dark ages and perspectives for SKA}

\noindent

Streaming motions affect in the first place the primordial gas evolution, delaying its collapse and the subsequent star formation history.
This is shown by {\it ad hoc} numerical simulations \cite[][]{Maio2011a} adopting a \lcdm{} model with present-day geometrical parameters for total-matter, $\Lambda$, dark-matter and baryon density given by:
$\omm = 0.3$,
$\omlambda = 0.7$,
$\omdm = 0.26$,
$\omb = 0.04$,
and an expansion parameter at the present $H_0 = 100 h \,\rm km/s/Mpc$ with $h=0.7$.
A simple way to take into account primordial bulk velocities in numerical simulations is to assume a velocity shift along the e.g. $x$-direction, $\vbx$, given uniformly to each gas particle at initial time, according to magnitudes of $\vbx = $ 0, 30, 60 km/s at $z\simeq 1020$ -- see \cite[][]{Naoz2011} for further studies.
In this way we explore various streaming velocity scenarios by means of the numerical code Gadget \cite[][]{Springel2005}, with subsequent modifications including: non-equilibrium cosmic chemistry (for e$^-$, H, H$^+$, H$^-$, He, He$^+$, He$^{++}$, H$_2$, H$^+_2$, D, D$^+$, HD, and HeH$^+$); stellar evolution of stars for different masses, metal yields (He, C, O, Mg, S, Si, Fe, N, etc.) and lifetimes \cite[][]{Maio2007, Tornatore2007, Maio2010}; star formation led by H, He and H$_2$ radiative losses in pristine environments and by resonant or fine-structure metal transitions (such as 
CII 157.7 $\mu$m, 
SiII 34.8 $\mu$m, 
OI 145.5 and 63.18 $\mu$m, 
FeII 87.41, 51.28, 35.35, 25.99 $\mu$m) in enriched environments \cite[][]{Maio2007}.
Subsequent supernova and wind feedback are responsible for spreading heavy elements in the surrounding gas and for determining the transition from the primordial pristine (population III) star formation regime to the metal-enriched solar-like (population II-I) regime when a critical metallicity of $10^{-4}\zsun$ is reached -- see further details in \cite{Maio2010}.
Cosmic structures at different redshifts are identified by means of a friends-of-friends algorithm with a linking length of 20\% the mean interparticle separation.
For each object gas, dark and stellar properties are stored.

\subsection{Implications for gas evolution}
\begin{figure*}
\centering
\includegraphics[width=0.45\textwidth]{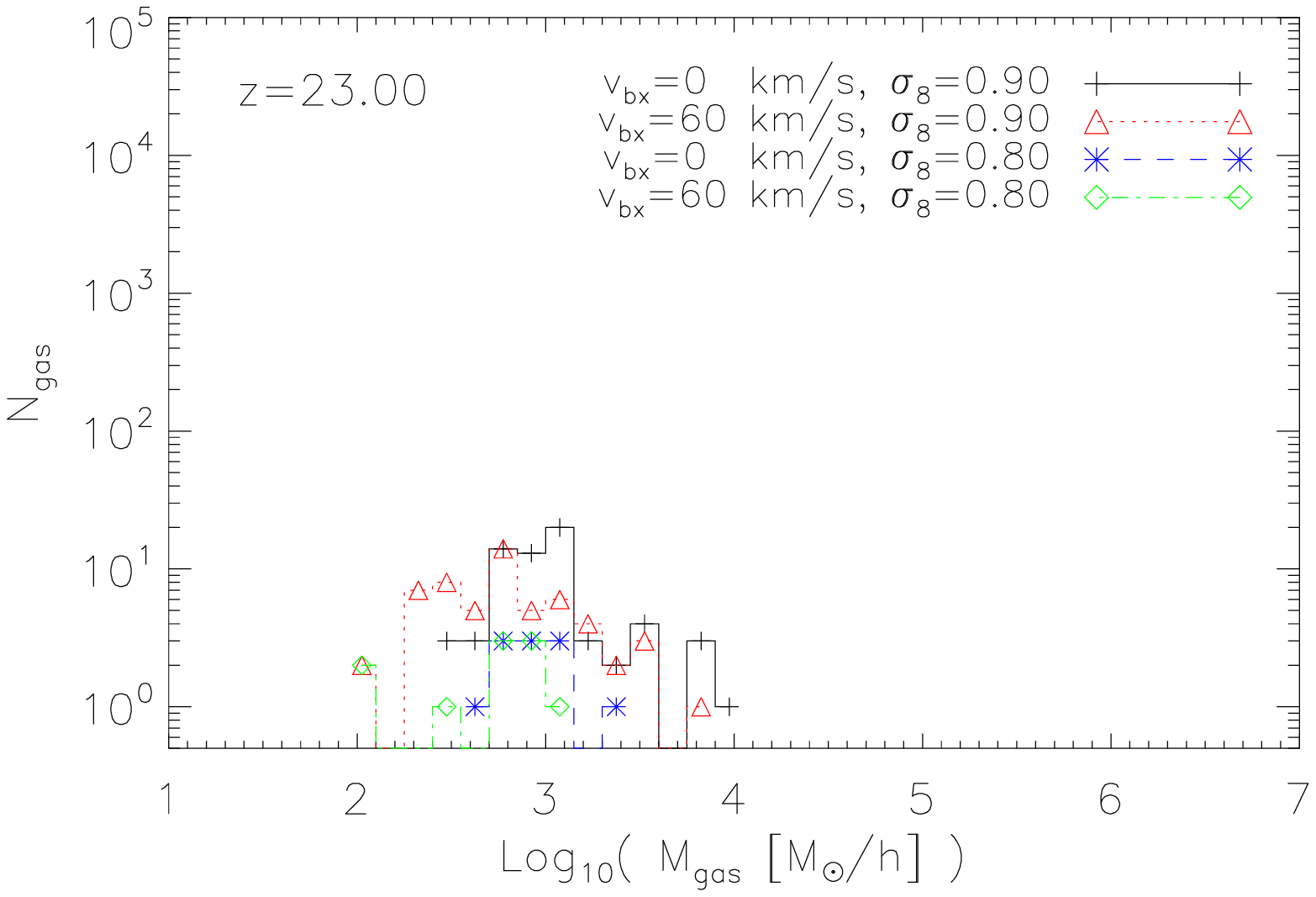}
\includegraphics[width=0.45\textwidth]{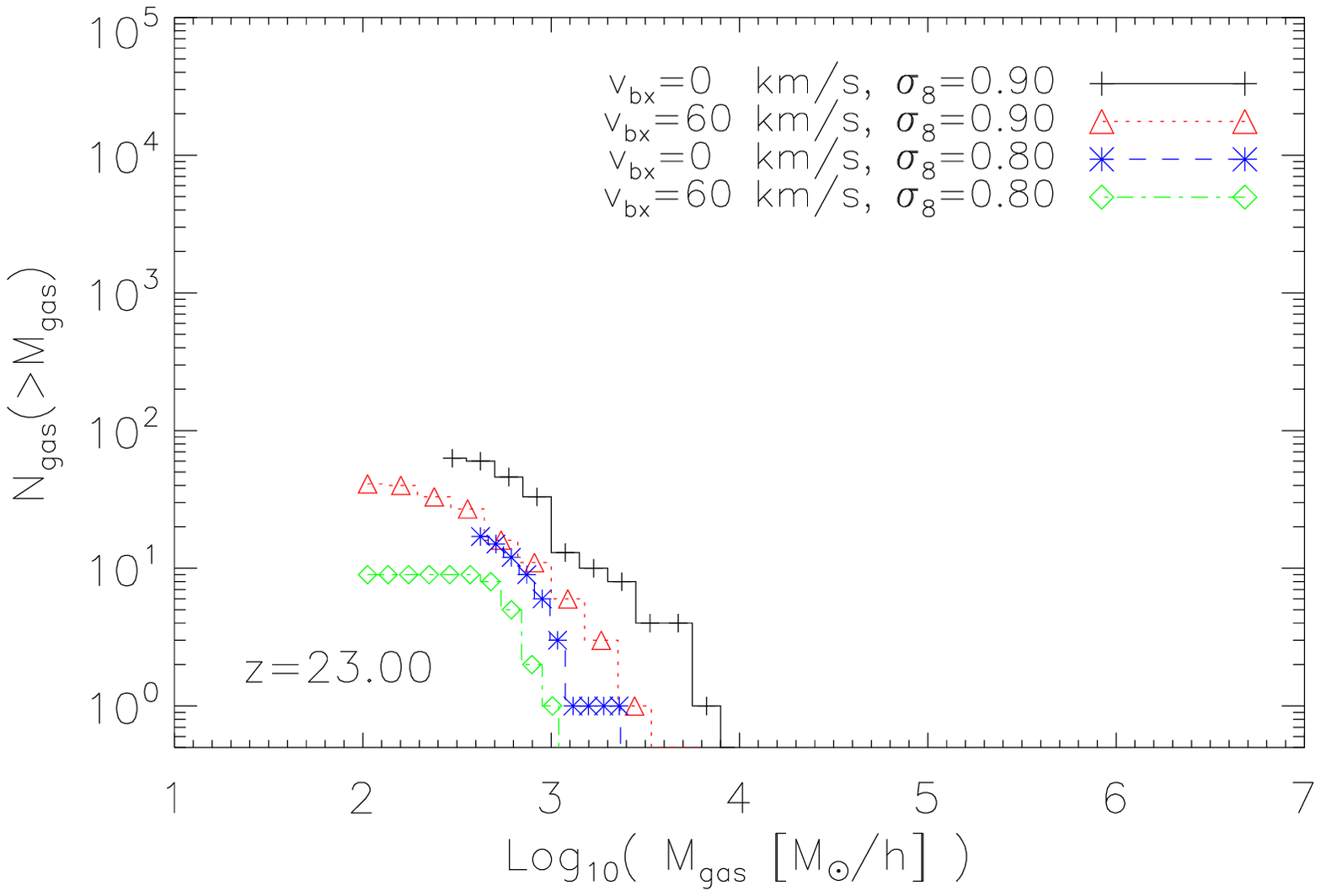}\\
\includegraphics[width=0.45\textwidth]{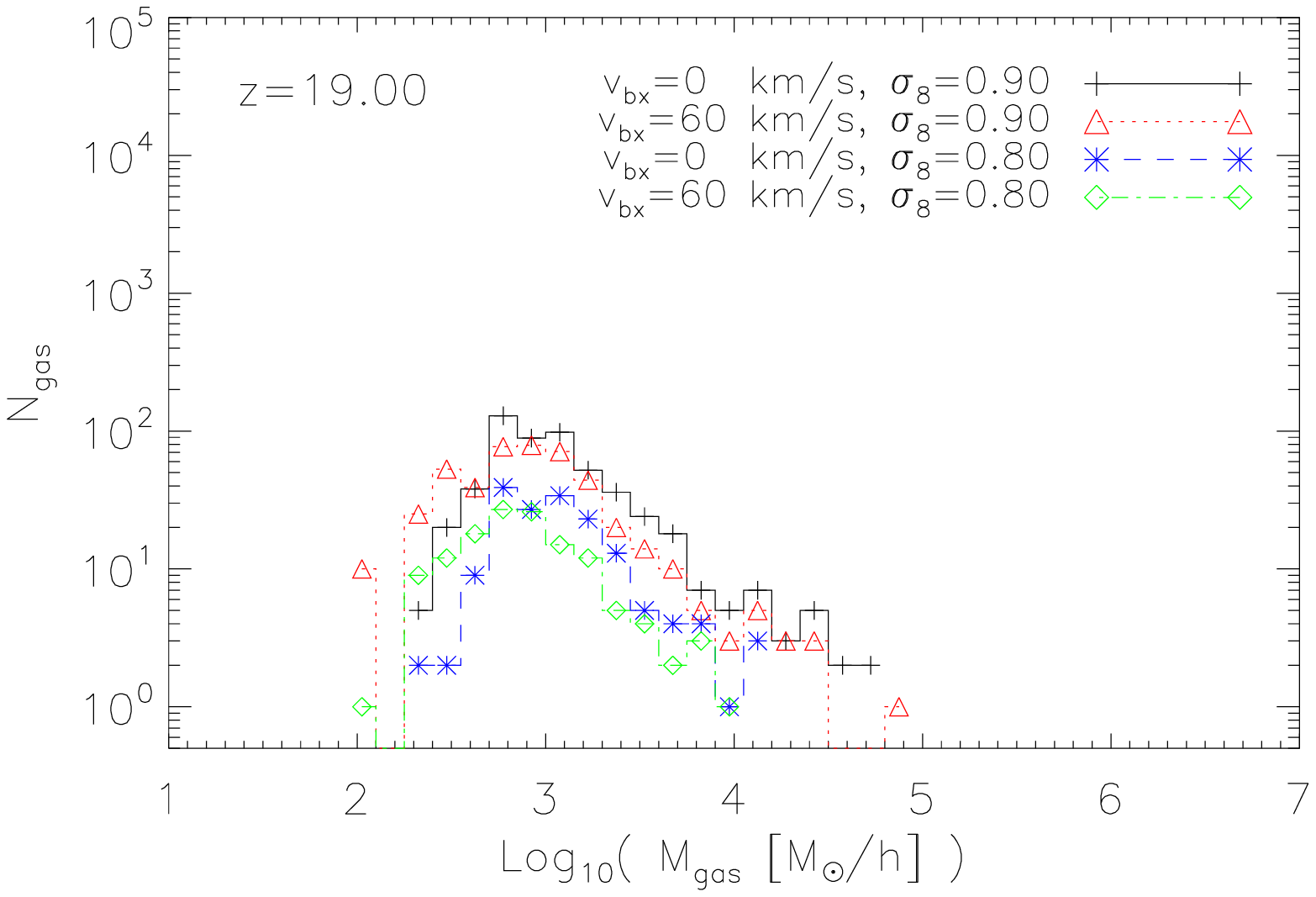}
\includegraphics[width=0.45\textwidth]{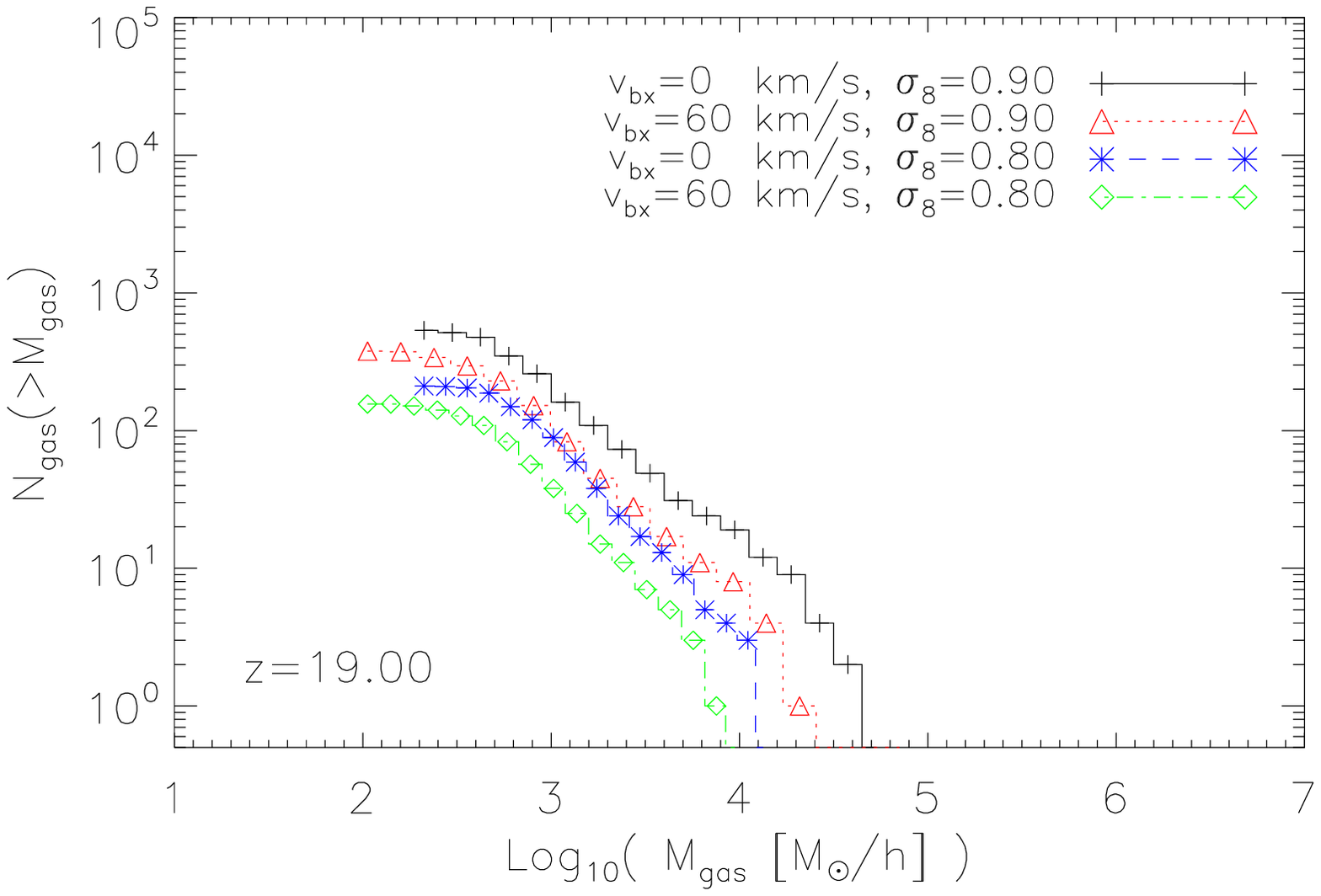}\\
\caption[]{\small
Expected differential (left column) and cumulative (right column) distributions for gas clouds at redshift $z=23$ (top) and $19$ (bottom).
Data refer to simulations of $\sim$1~Mpc-size box run with different $\sigma_8$ and $\vbx$:
$\sigma_8=0.9$ and $\vbx=0\,\rm km/s$ (solid line and crosses);
$\sigma_8=0.9$ and $\vbx=60\,\rm km/s$ (dotted line and triangles);
$\sigma_8=0.8$ and $\vbx=0\,\rm km/s$ (dashed line and asterisks);
$\sigma_8=0.8$ and $\vbx=60\,\rm km/s$ (dot-dashed line and squares).
}
\label{fig:distributions}
\end{figure*}

Fig.~\ref{fig:distributions} displays the differential (left) and cumulative (right) distribution of gas clouds at redshift $z=23$ (top) and $19$ (bottom), corresponding to cosmic times of about 0.1 and 0.2 Gyr, respectively, in different parameter scenarios.
They refer to four simulations run with initial conditions generated for a 0.5 Mpc/{\it h}-side box, sampling dark-matter and gaseous-matter fields with $2\times 320^3$ particles and corresponding to a gas resolution of about $40\,\msun$.
A primordial spectral slope of $n=1$ is assumed and the normalization is fixed by imposing $\sigma_8=0.9$.
In order to test degeneracies with spectral parameters, results for a case of $\sigma_8=0.8$ are overplotted, as well.
In general, streaming motions contribute in shaping the gaseous content and in lowering the typical clump masses and abundances.
This is mostly evident at higher redshift when dark-matter haloes are smaller and the additional kinetic energy coming from the bulk flows make more difficult to trap the gas in the dark-matter potential wells.
Panels referring to $z=23$ highlight this effect very clearly.
The differential distribution shows that, for any choice of $\sigma_8$, the $\vbx=0\,\rm km/s$ case predicts larger and more abundant gaseous structures, while the $\vbx=60\,\rm km/s$ scenario predicts typically lower-mass structures.
This is due to the effects of bulk flows that cause gas evacuation from primordial objects.
It is worth noting that the effects of streaming velocities is comparable to a significant decrease of $\sigma_8$ from 0.9 to 0.8.
The overall changes of cosmic abundances are better displayed by the cumulative distribution at the same redshift.
In the $\vbx=0\,\rm km/s$ case variations in the high-mass tail of the number densities of up to 1 dex with respect to the limiting case of $\vbx=60\,\rm km/s$ are reported.
In the lower-mass end suppression of a factor of $\sim 2$ is visible.
The slight excess of small clumps in the $\vbx>0\,\rm km/s$ models is due to residual gas that collapses later with respect to the $\vbx=0\,\rm km/s$ case.
At later times, it is still possible to see similar trends, as shown in the $z=19$ panels.
There differences between different-$\vbx$ models decrease slightly due to the simultaneous increase of typical dark-matter masses and the redshift decay of bulk flows.
Thus, gas in-fall and collapse are less hindered.
This leads to typical differences of a factor $\lesssim 2$ in the gas cloud differential distributions and of a factor of a few in the gas cloud cumulative distributions.
The statistical behaviour shows a converging evolution of the gas cloud masses and abundances at later times ($z\sim 10 - 20$).
At these epochs gas collapse is only marginally affected by primordial streaming motions (at $\sim 10\%$ level) and a major role is played by cosmological star formation and environments (see next).
These trends suggest that the main effect on gas cloud evolution is a delay of the collapse redshift of $\Delta z \sim $ a few (corresponding to tens of Myr at the epoch of interest).
\\
Since in these epochs there are no significant star formation events, yet, and forming structures are characterized by abundant pristine molecular gas, SKA could be able to detect H$_2$ and HD signatures from first collapsing sites.
E.g., H$_2$ $J=2-0$ rotational transition and HD $J=4-3$ rotational transition have rest-frame frequencies of $533.179\,\rm ~GHz$ and $533.388\,\rm ~ GHz$, respectively.
During Phase 1 SKA1-MID will have frequency bands up to $\sim $~14~GHz, hence observable (redshifted) signal from rare dense collapsing gas at very high redshift (around $z\sim 35-40$) will fall in the covered regime.
However, the SKA operational range will reach $24\,\rm ~GHz$ in phase 2, and then SKA frequency bands will cover the frequencies of molecular lines emitted by cold gas clouds down to $z\gtrsim 20$.
The emissivity of such lines should be detactable.
Indeed, H$_2$ $J=2-0$ emission is expected to be at level of $\sim 2\times 10^{-7}\,\rm Jy$ per each collapsing clump, while HD $J=4-3$ line is expected to have a lower rate of $\sim 10^{-8}\,\rm Jy$ \cite[][]{Kamaya2003}.
Roughly speaking, our theoretical estimates in Fig.~\ref{fig:distributions} suggest a number of collapsing clumps ranging from some tens up to some thousands per Mpc$^3$ at $z\sim 23-19$, hence the total emissivity by a $\sim 1 \,\rm Mpc^2$ patch (accounting for an area of $\vartheta\simeq 0.33\,\rm deg^2$ at $z\sim 19$)\footnote{
The (comoving) distance to $z\sim 19$ is about $10\,\rm Gpc$, the corresponding angular separation of 1 (comoving) Mpc results to be $10^{-2}\,\rm rad\simeq 0.573 \,\rm deg$ and hence $1 \,\rm Mpc^2$ patch covers $ 0.33\,\rm deg^2$. This is comparable to SKA1-MID resolution.
}
should range from $\sim 10$ up to $\sim 1000$ times $ 2\times 10^{-7}\,\rm Jy$, i.e. $\sim 2-200\,\rm \mu Jy$.
Assuming field of view and sensitivity for SKA1-MID ($\Delta = 0.49\,\rm deg^2$ and $63 \,\rm \mu Jy-hr^{-1/2}$, respectively) the detectable emission turns out to be $\Delta / \vartheta  = 0.49 / 0.33 \simeq 1.5$ times higher, i.e. around $\sim 3-300\,\rm \mu Jy$.
Therefore, the effects could be assessed by a-few-hours SKA1 observations of gas molecular lines.
The signal should be more easily detected by SKA2-LOW that is going to have a 10-times better sensitivity in the $350-24000\,\rm MHz$ frequency band.

\subsection{Implications for star formation}
\begin{figure}
\centering
\includegraphics[width=0.48\textwidth]{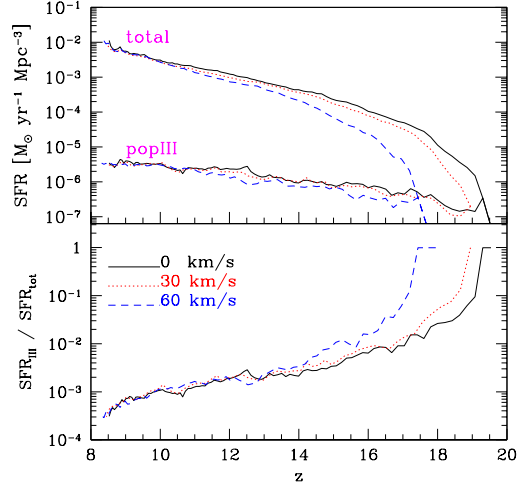}
\caption[]{\small
{\it Top panel:}
Redshift evolution of the total cosmological star formation rate density and of the population III (pop III) star formation rate density for
$\vbx =  0\,\rm km/s$ (solid lines),
$\vbx = 30\,\rm km/s$ (dotted lines),
$\vbx = 60\,\rm km/s$ (dashed lines).
{\it Bottom panel:} Contribution of the pop III star formation regime ($\rm SFR_{III} $) to the cosmological star formation rate ($\rm SFR_{tot}$) for the different $\vbx$ cases.
}
\label{fig:sfr}
\end{figure}
After the condensation of first gas clumps, star formation processes start taking place and radiation from stars act on the thermodynamical state of the surrounding medium.
\\
The top panel of Fig.~\ref{fig:sfr} displays the effects of streaming motions on the whole cosmological star formation rate density and on the population III (pop III) contribution.
In all the cases there is a cosmological effect on the trend and magnitude of the star formation activity.
This suffers the delays in the gas collapse caused by streaming motions and shows a retarded onset  for different $\vbx$ values, up to $\Delta z \sim 3$.
For the $\vbx = 0 \,\rm km/s$ case the onset takes place at $z\simeq 20$, when the age of the Universe is about $ 0.17\,\rm Gyr$, while for the cases including a velocity shift star formation kicks in at $z\simeq 19$ and $z\simeq 17.6$ for $\vbx = 30 \,\rm km/s$ and $\vbx = 60 \,\rm km/s$, respectively.
The corresponding cosmological times are $ 0.19\,\rm Gyr$, and $ 0.21\,\rm Gyr$ accounting for time delays of some ten Myr with respect to the $\vbx = 0 \,\rm km/s$ case.
Values for the star formation rate density differ at very early times of more than one order of magnitude and global trends show a converging behaviour only below $z\sim 14$, catching up at $z\lesssim 10$.
The reason for this is related to gas cloud evolution that results delayed because of the higher kinetic energy gained at recombination.
In larger-$\vbx$ models, primordial mini-haloes can retain less material (see Fig.~\ref{fig:distributions}), the hosted gas condenses more slowly, molecule formation gets hindered, the consequent cooling is less efficient and the resulting star formation activity is delayed.
This impacts mainly primordial haloes with masses $\sim 10^4$ -- $10^8\msun$, as their dimensions are comparable to or smaller than the baryon Jeans length \cite[][]{Tseliakhovich2010}.
Therefore, gas cannot fragment within the halo and partially or entirely flows out of it, as larger dark-matter potential wells would be needed to completely retain it.
Basically, these processes start from small scales and cascade over the growth of larger objects suppressing gas condensation and subsequent star formation already before reionization.
Photon production, feedback effects and the consequent occurrence of the first heavy elements in the Universe will be delayed, as well.
This is better seen in the lower panel of Fig.~\ref{fig:sfr} that highlights the contribution from the pristine pop III regime (whose evolution is dictated by ongoing metal production) to the cosmic star formation rate.
It is evident that for larger $\vbx$ the transition from the primordial pop III era to the metal-enriched pop II-I one takes place at later times due to the impacts of early bulk flows.
At later times ($z\lesssim 10$) the trends converge and the imprints of the initial dynamical patterns are more difficult to recognize.

\subsection{Implications for 21-cm signal}   

A valuable observable to investigate the cosmological response of baryons at the end of the dark ages ($z\sim 6-20$) is 21-cm signal from HI.
This falls in the frequency range covered by SKA1-LOW ($50-350$~MHz), that could possibly map at a 3$-\sigma$ level neutral gas from the dark ages and cosmic dawn until full re-ionization at $z\sim 6$ \cite[][]{Mellema2013}.
\\
The expected signal is usually described in terms of differential brightness temperature, $\delta T_b$, the extent of which depends mainly on the ionization and thermodynamical state of the (partially) neutral gas.
The effect on star formation caused by primordial streaming motions will impact $\delta T_b$ in different ways. In particular, by delaying the onset of star formation streaming motions would also shift the beginning of the re-ionization process towards lower redshift (i.e. higher frequencies), while by suppressing or reducing star formation within small mass halos they would change the topology of re-ionization.
As a consequence, an impact on the 21-cm signal is also expected.
Semi-analytical arguments extrapolated to $\sim 100\,\rm Mpc$ scales \cite[][]{Dalal2010,Visbal2012,McQuinn2012,Ali2014} suggest $\vbx $ implications on the 21-cm signal from comoving distances of the order of $\sim 100 \,\rm kpc$ \cite[][]{Visbal2012,McQuinn2012} down to $\sim 10 \,\rm kpc$ \cite[][]{McQuinn2012, Ali2014}, with an overall effect on $\delta T_{\rm b}$ ranging from $\sim  2\,\rm mK$ \cite[][]{Dalal2010} to $ \sim 10\,\rm mK$ \cite[][]{Visbal2012}.
On the other hand, implications from structure formation shocks can contribute at $\lesssim 10\%$ level, while enhancements to the ionization fraction due to X-rays, albeit very uncertain, seem to be negligible or almost irrelevant \cite[e.g.][]{McQuinn2012}.
\\
We finally note that the scenario commonly adopted for these estimates relies on standard cold dark matter, however, comparable or even larger suppressions of cosmic structure evolution might arise in presence of alternative background models, such as warm dark matter \cite[see e.g.][and references therein]{MaioViel2014}, non-Gaussian matter distributions \cite[][]{MaioIannuzzi2011, Maio2012nG, MaioKhochfar2012, Pace2014} or early dark energy \cite[][]{Maio2006}.

\section{Conclusions}
Higher-order corrections \cite[][]{Tseliakhovich2010} to the cosmological linear perturbation theory predict the formation of coherent supersonic gaseous streaming motions at decoupling.
These are relevant for primordial gas condensation, the birth of the first stars and the build-up of the lowest-mass objects and proto-galaxies.
In view of the upcoming SKA, we have discussed the implications of such bulk flows via numerical N-body hydrodynamical chemistry simulations \cite[][]{Maio2011a} of early structure formation taking into account non-equilibrium cosmic chemistry, gas cooling \cite[][]{Maio2007}, stellar evolution of stars for different masses, metal yields (He, C, O, Mg, S, Si, Fe, N, etc.) and lifetimes \cite[][]{Maio2007, Tornatore2007} and the transition \cite[][]{Maio2010} from the primordial pristine pop III star formation regime to the metal-enriched solar-like pop II-I.
Primordial gas streaming motions can suppress gas collapse, with consequent delay of cosmological star formation, feedback effects, metal spreading and cosmic re-ionization.
Their impacts on gas clouds are particularly evident at $z\sim 15-30$, when number densities and typical gas content in primordial haloes result lowered of up to a factor of 10.
Hence, they affect molecular H$_2$ and/or HD emissions from overdense gaseous regions that will be observable by SKA1-MID.
Subsequently, bulk flows delay star formation, metal pollution and feedback mechanisms of $\Delta z\sim$ a few at $z\sim 10-20$, influencing the cosmic HI thermal and chemical state, and making the 21-cm signal at the end of the dark ages patchier and noisier.
The next SKA mission will represent a major advance over existing instruments, since it will be able to detect HI 21-cm (e.g. SKA1-LOW at $ 50-350 \, \rm MHz$) and molecular-line (e.g. SKA1-MID up to $14-24\,\rm GHz$ in phase 1 -- 2) emissions from early star forming sites at $z\sim 6 - 40$ (SKA2-LOW).
Therefore, it will be possible to address the effects of primordial streaming motions on early baryon evolution and to give constraints on structure formation in the first Gyr.


\bibliographystyle{apalike}


\end{document}